\documentclass[journal]{IEEEtran}
\usepackage[dvipsnames]{xcolor}
\usepackage{tikz} 
\usepackage{amsmath,amsfonts}
\usepackage{algorithmic}
\usepackage{algorithm}
\usepackage{array}
\usepackage{textcomp}
\usepackage{stfloats}
\usepackage{url}
\usepackage{verbatim}
\usepackage{graphicx}
\usepackage{cite}
\usepackage{pdfpages}
\usepackage{tikz,mathpazo}  
\usetikzlibrary{shadows,arrows,positioning,shapes.geometric,chains} 
\usetikzlibrary{calc} 
\tikzstyle{straightline} = [line width = 1pt,-]
\usepackage{caption}
\usepackage{subcaption}
\usepackage{authblk}
\usepackage{multirow,booktabs,siunitx} 
\ifCLASSINFOpdf
\else
   \usepackage[dvips]{graphicx}
\fi
\usepackage{url}

\usepackage{graphicx}

\begin{document}

\title{Fast and Blind Speech Copy-Move Detection and Localization in Noise}

\author{Dong~Yang,~Mingle~Liu,~Muyong~Cao
\thanks{Dong Yang, Mingle Liu, and Muyong Cao are with the GVoice from IEG, Tencent Holdings Ltd. (e-mail:d.yang.be@gmail.com ) }
}
\markboth{Journal of \LaTeX\ Class Files, Vol. 14, No. 8, August 2015}
{Shell \MakeLowercase{\textit{et al.}}: Bare Demo of IEEEtran.cls for IEEE Journals}
\maketitle
\begin{abstract}
Copy-move forgery on speech (CMF), coupled with post-processing techniques, presents a great challenge to the forensic detection and localization of tampered areas. Most of the existing CMF detection approaches necessitate pre-segmentation of speech to facilitate similarity calculations among these segments. However, these approaches usually suffer from the problems of uncontrollable computational complexity and sensitivity to the presence of a word that is read multiple times within a speech recording. To address these issues, we propose a local feature tensors-based CMF detection algorithm that can transform duplicate detection and localization problems into a special tensor-matching procedure, accompanied by complete theoretical analysis as support. Through extensive experimentation, we have demonstrated that our method exhibits computational efficiency and robustness against post-processing techniques. Notably, it can effectively and blindly detect tampered segments, even those as short as a fractional second. These advantages highlight the promising potential of our approach for practical applications.
\end{abstract}

\begin{IEEEkeywords}
Speech forensics, blind detection, copy-move detection, local feature tensors
\end{IEEEkeywords}

\IEEEpeerreviewmaketitle

\section{Introduction}

\IEEEPARstart{C}{opy-move} forgery speech is often imperceptible to human beings due to its derivation from real prerecorded recordings and manipulation using sophisticated audio editing tools. Furthermore, speech recordings in everyday life can be lengthy, while the tampered parts within a recording may be sparse in time. Hence, detecting CMF through auditory and visual analysis becomes challenging. Additionally, post-processing manipulations, including filtering, compression, resampling, and even the presence of background noise and music, further complicate the detection process.

In recent years, scholars have shown a growing interest in speech forensics, resulting in numerous reported works in this field\cite{Bevinamarad2020AudioFD}. Most of the existing forgery detection techniques can be broadly classified into active and passive techniques. Passive techniques \cite{zhao2014audio} \cite{imran2017blind} are considered more practical because they verify the authenticity of audio by analyzing its contents and structure. Q. Yans et al \cite{yan2019robust} \cite{Yan2017} proposed an acoustic features similarities-based CMFs detecting method. F. Wang et al \cite{wang2017algorithm} presented discrete cosine transforms and singular value decomposition techniques. Z. Liu et al \cite{Liu2017FastCD} discussed the Pearson correlation coefficient (PCC) based method in the discrete Fourier transform (DFT) domain to verify the similarity of the speech segments and M. Imran et al \cite{imran2017blind} compare the histograms computed via 1-D local binary pattern operator to identify locations of CMF. The main drawbacks of the existing CMF detection methods can be summarized as follows:

1) These methods usually assume that the speech recording has sufficient between-utterance silence \cite{imran2017blind}\cite{yan2019robust}\cite{Akdeniz2022}\cite{Yan2017}\cite{XIE201837}. However, in real conversation scenarios, it is evident that this assumption does not hold, and thus they are not able to perform a fully blind analysis.

2) These methods rely on similarity computation\cite{yan2019robust} \cite{Yan2017}, which leads to the need for complex thresholds tuning and tends to misjudge when a word is read multiple times within a speech recording. Additionally, their computational complexity is non-linear, making the computation time unpredictable for long recordings.

3) These methods can work well in speech recordings without or with only light post-processing, but they may be sensitive to severe post-processing attacks.

In this work, an efficient, fully blind, and robust speech anti-CMF algorithm based on local feature tensors (LFTs) that transforms duplicate detection and localization problems into a special tensor-matching procedure is proposed. The rest of this paper is organized as follows. In Section II, we first revisit the issues of speech CMF and then show the details of our proposed scheme. The experimental results are presented in Section III, followed by a conclusion in Section IV.

\section{APPROACH}
\subsection{Framework}
\begin{figure}
\centering
\scalebox{0.50}{\tikzset{every picture/.style={line width=0.75pt}} 

\begin{tikzpicture}[x=0.75pt,y=0.75pt,yscale=-1,xscale=1]

\draw  [fill={rgb, 255:red, 255; green, 253; blue, 249 }  ,fill opacity=1 ][dash pattern={on 4.5pt off 4.5pt}] (203.99,40.6) .. controls (203.99,20.39) and (220.38,4) .. (240.59,4) -- (606.74,4) .. controls (626.95,4) and (643.34,20.39) .. (643.34,40.6) -- (643.34,150.4) .. controls (643.34,170.61) and (626.95,187) .. (606.74,187) -- (240.59,187) .. controls (220.38,187) and (203.99,170.61) .. (203.99,150.4) -- cycle ;
\draw  [fill={rgb, 255:red, 255; green, 184; blue, 184 }  ,fill opacity=1 ] (24.12,11) .. controls (24.12,6.58) and (27.7,3) .. (32.12,3) -- (151.95,3) .. controls (156.37,3) and (159.95,6.58) .. (159.95,11) -- (159.95,35) .. controls (159.95,39.42) and (156.37,43) .. (151.95,43) -- (32.12,43) .. controls (27.7,43) and (24.12,39.42) .. (24.12,35) -- cycle ;
\draw  [fill={rgb, 255:red, 255; green, 184; blue, 184 }  ,fill opacity=1 ] (4,82) .. controls (4,77.58) and (7.58,74) .. (12,74) -- (169.84,74) .. controls (174.26,74) and (177.84,77.58) .. (177.84,82) -- (177.84,106) .. controls (177.84,110.42) and (174.26,114) .. (169.84,114) -- (12,114) .. controls (7.58,114) and (4,110.42) .. (4,106) -- cycle ;
\draw  [fill={rgb, 255:red, 248; green, 231; blue, 28 }  ,fill opacity=1 ] (455.08,82) .. controls (455.08,77.58) and (458.67,74) .. (463.08,74) -- (631.54,74) .. controls (635.96,74) and (639.54,77.58) .. (639.54,82) -- (639.54,106) .. controls (639.54,110.42) and (635.96,114) .. (631.54,114) -- (463.08,114) .. controls (458.67,114) and (455.08,110.42) .. (455.08,106) -- cycle ;
\draw  [fill={rgb, 255:red, 213; green, 178; blue, 253 }  ,fill opacity=1 ] (225.91,82) .. controls (225.91,77.58) and (229.49,74) .. (233.91,74) -- (402.37,74) .. controls (406.78,74) and (410.37,77.58) .. (410.37,82) -- (410.37,106) .. controls (410.37,110.42) and (406.78,114) .. (402.37,114) -- (233.91,114) .. controls (229.49,114) and (225.91,110.42) .. (225.91,106) -- cycle ;
\draw  [fill={rgb, 255:red, 231; green, 255; blue, 207 }  ,fill opacity=1 ] (336.58,15) .. controls (336.58,10.58) and (340.16,7) .. (344.58,7) -- (513.04,7) .. controls (517.46,7) and (521.04,10.58) .. (521.04,15) -- (521.04,39) .. controls (521.04,43.42) and (517.46,47) .. (513.04,47) -- (344.58,47) .. controls (340.16,47) and (336.58,43.42) .. (336.58,39) -- cycle ;
\draw  [fill={rgb, 255:red, 80; green, 227; blue, 194 }  ,fill opacity=1 ] (337.7,151) .. controls (337.7,146.58) and (341.28,143) .. (345.7,143) -- (514.16,143) .. controls (518.58,143) and (522.16,146.58) .. (522.16,151) -- (522.16,175) .. controls (522.16,179.42) and (518.58,183) .. (514.16,183) -- (345.7,183) .. controls (341.28,183) and (337.7,179.42) .. (337.7,175) -- cycle ;
\draw    (195.5,163) -- (332.35,163) ;
\draw [shift={(334.35,163)}, rotate = 180] [color={rgb, 255:red, 0; green, 0; blue, 0 }  ][line width=0.75]    (10.93,-3.29) .. controls (6.95,-1.4) and (3.31,-0.3) .. (0,0) .. controls (3.31,0.3) and (6.95,1.4) .. (10.93,3.29)   ;
\draw  [fill={rgb, 255:red, 255; green, 184; blue, 184 }  ,fill opacity=1 ] (4,151) .. controls (4,146.58) and (7.58,143) .. (12,143) -- (169.84,143) .. controls (174.26,143) and (177.84,146.58) .. (177.84,151) -- (177.84,175) .. controls (177.84,179.42) and (174.26,183) .. (169.84,183) -- (12,183) .. controls (7.58,183) and (4,179.42) .. (4,175) -- cycle ;
\draw    (315.9,74) -- (315.9,29) -- (335.7,29) ;
\draw [shift={(337.7,29)}, rotate = 180] [color={rgb, 255:red, 0; green, 0; blue, 0 }  ][line width=0.75]    (10.93,-3.29) .. controls (6.95,-1.4) and (3.31,-0.3) .. (0,0) .. controls (3.31,0.3) and (6.95,1.4) .. (10.93,3.29)   ;
\draw    (490.3,47) -- (490.3,71) ;
\draw [shift={(490.3,73)}, rotate = 270] [color={rgb, 255:red, 0; green, 0; blue, 0 }  ][line width=0.75]    (10.93,-3.29) .. controls (6.95,-1.4) and (3.31,-0.3) .. (0,0) .. controls (3.31,0.3) and (6.95,1.4) .. (10.93,3.29)   ;
\draw    (489.18,115) -- (489.18,140) ;
\draw [shift={(489.18,142)}, rotate = 270] [color={rgb, 255:red, 0; green, 0; blue, 0 }  ][line width=0.75]    (10.93,-3.29) .. controls (6.95,-1.4) and (3.31,-0.3) .. (0,0) .. controls (3.31,0.3) and (6.95,1.4) .. (10.93,3.29)   ;
\draw    (522.72,163) -- (656,163) ;
\draw [shift={(658,163)}, rotate = 180] [color={rgb, 255:red, 0; green, 0; blue, 0 }  ][line width=0.75]    (10.93,-3.29) .. controls (6.95,-1.4) and (3.31,-0.3) .. (0,0) .. controls (3.31,0.3) and (6.95,1.4) .. (10.93,3.29)   ;
\draw    (91.2,43) -- (91.2,73) ;
\draw [shift={(91.2,75)}, rotate = 270] [color={rgb, 255:red, 0; green, 0; blue, 0 }  ][line width=0.75]    (10.93,-3.29) .. controls (6.95,-1.4) and (3.31,-0.3) .. (0,0) .. controls (3.31,0.3) and (6.95,1.4) .. (10.93,3.29)   ;
\draw    (92.32,115) -- (92.32,140) ;
\draw [shift={(92.32,142)}, rotate = 270] [color={rgb, 255:red, 0; green, 0; blue, 0 }  ][line width=0.75]    (10.93,-3.29) .. controls (6.95,-1.4) and (3.31,-0.3) .. (0,0) .. controls (3.31,0.3) and (6.95,1.4) .. (10.93,3.29)   ;
\draw    (185.84,163) -- (185.84,95) -- (221.5,95) ;
\draw [shift={(223.5,95)}, rotate = 180] [color={rgb, 255:red, 0; green, 0; blue, 0 }  ][line width=0.75]    (10.93,-3.29) .. controls (6.95,-1.4) and (3.31,-0.3) .. (0,0) .. controls (3.31,0.3) and (6.95,1.4) .. (10.93,3.29)   ;
\draw    (178,163) -- (185.84,163) ;
\draw    (160.5,24) -- (195.5,24) ;
\draw    (195.5,24) -- (195.5,68) -- (195.5,163) ;

\draw (50.05,13) node [anchor=north west][inner sep=0.75pt]   [align=left] {Audio wave$ $};
\draw (46.67,86) node [anchor=north west][inner sep=0.75pt]   [align=left] {Preprecessing};
\draw (30.34,147) node [anchor=north west][inner sep=0.75pt]  [font=\small] [align=left] {\begin{minipage}[lt]{79.44pt}\setlength\topsep{0pt}
\begin{center}
Logarithmic scale \\STFT spectogram
\end{center}

\end{minipage}};
\draw (371.75,10) node [anchor=north west][inner sep=0.75pt]  [font=\small] [align=left] {\begin{minipage}[lt]{77.86pt}\setlength\topsep{0pt}
\begin{center}
Local feature\\tensors extracting
\end{center}

\end{minipage}};
\draw (230.14,75) node [anchor=north west][inner sep=0.75pt]   [align=left] {\begin{minipage}[lt]{124.95pt}\setlength\topsep{0pt}
\begin{center}
{\small Harmonic Constellation map }\\{\small analysis}
\end{center}

\end{minipage}};
\draw (466.19,84) node [anchor=north west][inner sep=0.75pt]   [align=left] {\begin{minipage}[lt]{113.33pt}\setlength\topsep{0pt}
\begin{center}
{\small Fast duplication searching}
\end{center}

\end{minipage}};
\draw (339.7,154) node [anchor=north west][inner sep=0.75pt]   [align=left] {\begin{minipage}[lt]{123.38pt}\setlength\topsep{0pt}
\begin{center}
{\small Local pitch distance analysis}
\end{center}

\end{minipage}};
\draw (601.86,142) node [anchor=north west][inner sep=0.75pt]  [font=\small] [align=left] {\begin{minipage}[lt]{29.2pt}\setlength\topsep{0pt}
\begin{center}
output
\end{center}

\end{minipage}};
\draw (172,123.4) node [anchor=north west][inner sep=0.75pt]    {$\mathbf{S}$};
\draw (320,47.4) node [anchor=north west][inner sep=0.75pt]    {$\mathbf{P}$};
\draw (502,50.4) node [anchor=north west][inner sep=0.75pt]    {$\mathbf{T}$};
\draw (552,140.4) node [anchor=north west][inner sep=0.75pt]    {$\mathbf{M}^{f}$};
\draw (174,6.4) node [anchor=north west][inner sep=0.75pt]    {$\mathbf{x}$};
\draw (502,120.4) node [anchor=north west][inner sep=0.75pt]    {$\mathbf{M}$};

\end{tikzpicture}}
\caption{Flowchart of the proposed approach}
\label{fig:flowchat}
\end{figure}
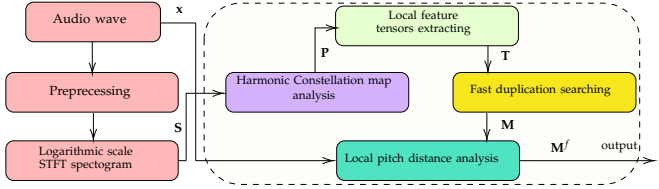

 Fig.~\ref{fig:flowchat} presents the framework of our proposed speech forgery detection scheme. The search and localization of duplicated segments in the speech recordings are computed in the logarithmic STFT magnitude (LSTFTM) spectrum domain by exploring the attributes of time–frequency~(T-F) spectrogram representation, namely the LFTs. The principles for designing the LFTs are that they should be temporally and frequentially localized, temporally translation invariant, robust to channel interference, and sufficiently entropic. We obtain the LFTs from the harmonic constellation map (HCM) on a speech spectrogram, which can be readily incorporated into CMF identification and localization implemented by an efficient hash-based duplication searching algorithm. In the following paragraphs, we will introduce the process of the proposed method in detail.
\subsection{Harmonic constellation map analysis (HCMA) }
\begin{figure}
     \centering
     \begin{subfigure}[t]{0.5\textwidth}
         \centering
         \scalebox{0.50}{\input{./figures/harmonic_peaks}}
         \caption{}
         \label{fig:harmonic}
     \end{subfigure}
     \begin{subfigure}[t]{0.24\textwidth}
         \centering
         \includegraphics[width=\textwidth]{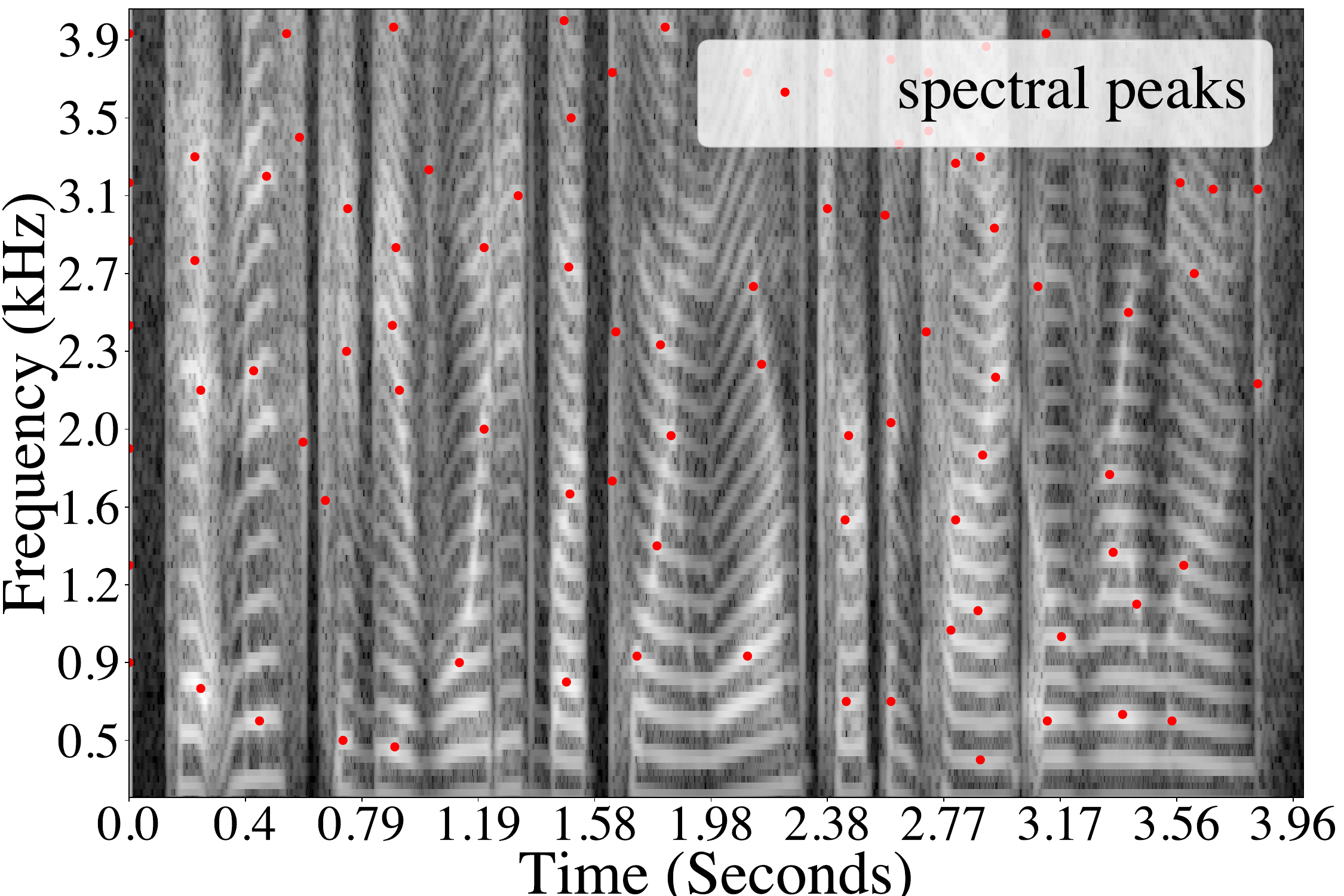}
         \caption{${\mathbf{B}_{d}}$}
         \label{fig:densekernel}
     \end{subfigure}
     \begin{subfigure}[t]{0.24\textwidth}
         \centering
         \includegraphics[width=\textwidth]{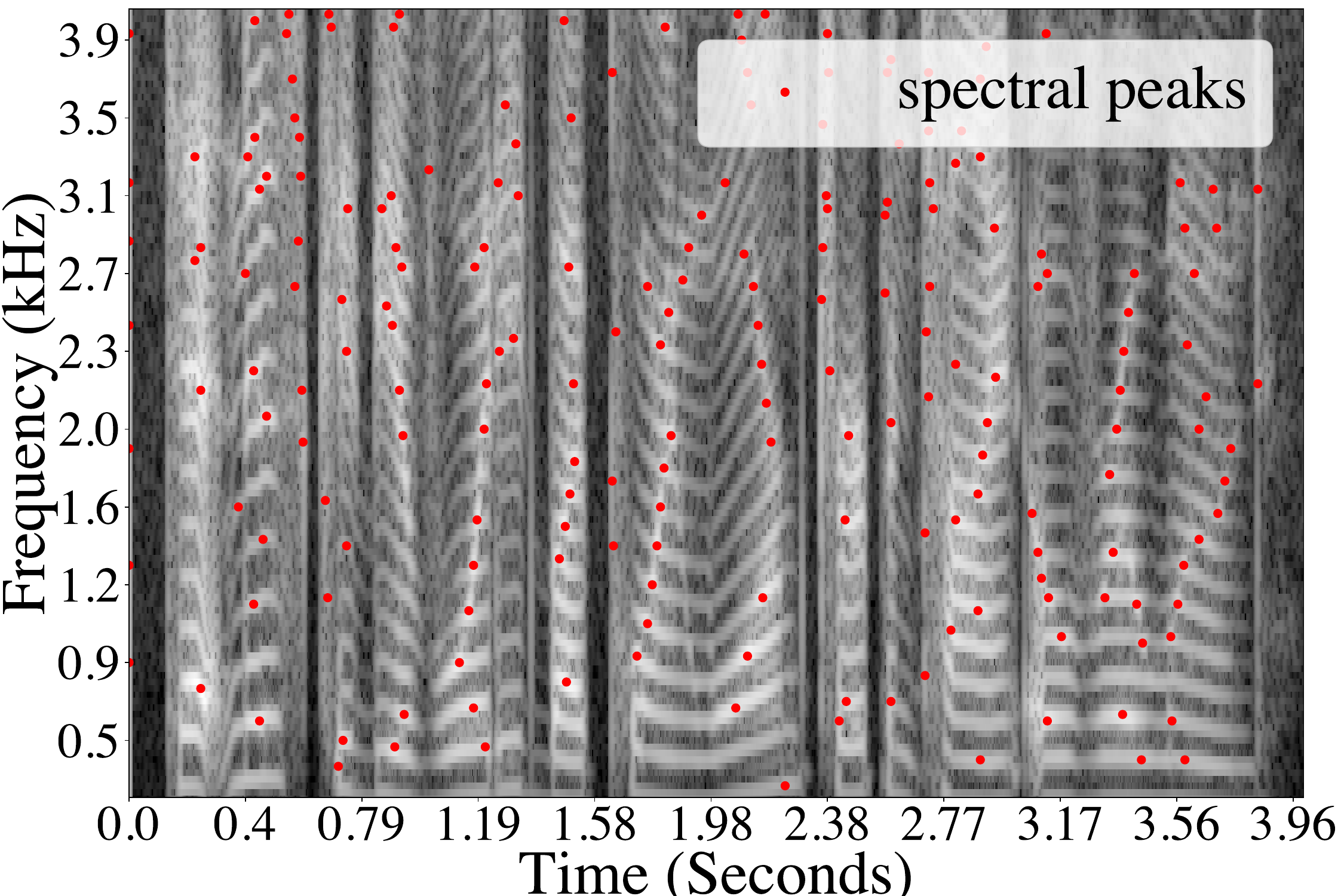}
         \caption{${\mathbf{B}_{s}}$}
         \label{fig:sparsekernel}
     \end{subfigure}
     \caption{(a) Comparison of the peak detection scheme with DNB and SNB on harmonics. (b) and (c) peak detection on the spectrogram with DNB and SNB respectively. }
\label{fig:peaksharmonic}
\end{figure}

LSTFTM is used as a quadratic T-F representation analysis tool in speech processing. To compensate for the spectral shape, a pre-emphasis filter is initially applied to emphasize higher frequencies of the speech signal $\mathbf{x}$. 
Finally, LSTFTM  can be expressed as
 \begin{equation}
 S(t, f)=20log_{10}\left| X(t, f) \right|=20log_{10}\left| {STFT}\left\{ \mathbf{x} \right\} \right|
 \label{eq:stft}
 \end{equation}
A point ${\left (i, j  \right )}$ on the spectrogram ${\mathbf{S}}\!=\!\{{S(t,f)}\}$ is considered a candidate peak if it has maximal amplitude among its neighbors in a region ${\mathbf{B}_{i,j}}$, centered around of it, with the shape of ${\mathbf{B}}$. All these identified peaks are denoted as ${\mathbf{P}}$ and can be calculated as follows:
 \begin{small}
\begin{equation}
  \mathbf{P}=\{(i,j) \mid S\left (i,j\right) \geq S\left( {i}', {j}' \right ), \forall ({i}',{j}') \in \mathbf{B}_{i,j}\}
\end{equation}
\end{small}
We choose spectrogram peaks as key points due to their robustness in the presence of background noise\cite{wang2003industrial}. However, selecting a dense neighborhood (DNB) ${\mathbf{B}}$, such as a rectangle shape  ${\mathbf{B}_d=[h,d]}$, leads to sparser and more scattered peaks, as illustrated in Fig \ref{fig:densekernel}. Considering the harmonic characteristics of voiced speech, opting for a sparse neighborhood (SNB) ${\mathbf{B}_s}$ yields a higher count of peaks along the harmonics curves on the spectrogram. This effect is particularly evident in the rapidly changing portion of the harmonics over time, as shown in Fig.~\ref{fig:sparsekernel}. Due to the typically higher transient speech energies along the harmonic curves, selecting an SNB increases robustness against noise. Moreover, the harmonics on the spectrogram exhibit smooth curves over time, resulting in a more structurally harmonic pattern of peaks. Consequently, it increases the presence of valid spectrogram peaks of speech. To effectively capture the energy variations of the harmonic curves on the spectrogram, we devise an SNB in the shape of a flat cross with parameters ${\mathbf{B}_s}$=${[{h_{1}}}$, ${{h_{2}}}$, ${{d}}$, ${d_{1}]}$. 
Fig.~\ref{fig:harmonic} presents a conceptual illustration and comparison of the peak detection scheme, showing the differences between the rectangle and flat cross neighborhoods. In the figure, the SNB, with an equivalent scale of the receptive field, demonstrates superior tracking of local peaks on harmonics compared to the DNB. This sparse approach also effectively avoids false peaks resulting from noise interference between harmonics. Among the parameters of the ${\mathbf{B}_s}$, the parameter ${{h_{1}}}$ controls the capturing of the magnitude information on the harmonics and regulates the sampling interval for peak detection along the time axis. Parameter ${{h_{2}}}$ controls the capturing of the magnitude information between adjacent harmonics along the frequency axis. Assuming we know the average pitch frequency ${{F_{0}}}$, sampling rate ${{R}}$ and FFT size ${L}$, a good choice for ${{h_{2}}}$ can be ${{2F_{0}L}/{R}}$.
\subsection{Local feature tensors extracting}
\label{sec:LFTE}
\begin{figure}
\scalebox{0.50}{\input{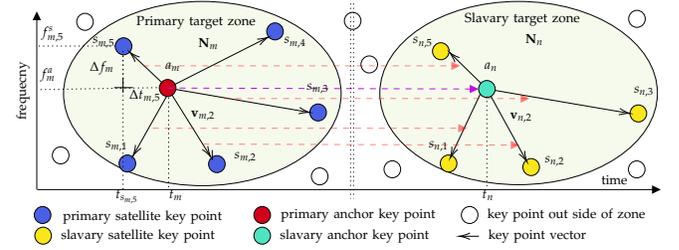}}
\caption{Illustration of LFTs extraction scheme and matching process of the LFTs pair.}
\label{fig:tensors}
\end{figure}

In this stage, LFTs of the speech signals are computed based on the HCMA. Initially, the ${m^{th}}$ anchor ${a_{m}=(f_{m}^a, t_m) \in \mathbf{P}}$ and a target zone ${\mathbf{N}_m}$ centered around ${a_{m}}$ are identified on the HCM. Then the anchor ${a_{m}}$ is sequentially paired with the ${l^{th}}$  satellite peak ${s_{m,l}=(f_{m,l}^s,t_{s_{m,l}})}$ within ${\mathbf{N}_m}$, as depicted in Fig.~\ref{fig:tensors}, yielding a vector consisting of two frequency components plus the time difference between them ${\Delta{t_{m,l}}}$, written as ${\mathbf{v}_{m,l}=[f_{m}^a \  ,\ f_{m,l}^s \ , \ \Delta{t_{m,l}}] }$, which is temporally translation invariant.
We define ${{\mathbf{V}}_m}=\{{\mathbf{v}}_{m,l}\}$, ${{\mathbf{V}}=\{{\mathbf{V}_{m}}\}}$, extended vector ${{\mathbf{v}_{m,l}^{e}}=[\mathbf{v}_{m,l} \,  \  t_{m}]}$, ${{\mathbf{V}_{m}^{e}}=\{{\mathbf{v}_{m,l}^{e}}\}}$ and ${{\mathbf{V}}^{e}=\{{\mathbf{V}_{m}^{e}}\}}$ that contains the absolute timestamps of the anchors so that we can later retain the duplicate time offset. It is assumed that fan-out number ${F_m}\!= \!\lvert{\mathbf{V}}_m\rvert$. To ensure manageable computational complexity, the number of elements of ${{\mathbf{V}}_m}$ is limited to $F$ by discarding the excess elements. ${f_m^a}$ and ${f_{m,l}^s}$ are integers less than half of FFT-size ${L/2}$, and the range of ${\Delta{t_{m,l}}}$ is limited in the target zone, denoted by ${N_{{x}}}$. Thus ${{\mathbf{v}}_{m,l}}$ can be embedded to an integer ${z_{m,l}}$ without information loss as follows:
\newcommand{\ceil}[1]{\lfloor #1 \rfloor}
\begin{equation}
\label{vec2integer}
z_{m,l}=2^{(L/2+\lceil{\log_2N_{{x}}}\rceil)}f_m^a+2^{(\lceil{\log_2N_{{x}}}\rceil)}{f_{m,l}^s}+\Delta{t_{m,l}}
\end{equation}
The count of ${z_{m,l}}$ equals to ${N_pF}$, where ${N_p}$ is the total number of peaks, which can be given by the density of peaks ${\kappa}$ times duration of recording  ${T_s}$. The value of $\kappa$ generally depends on the size and geometry of $\mathbf{B}$. In order to further reduce the computational complexity, we eliminate these candidates who have no duplication in ${{\mathbf{V}}}$ as they are not related to the copy-move operation. This is achieved in (\ref{delete}) by hash-based duplicate detection method \cite{leiserson1994introduction} on ${\mathbf{z}}$ that has complexity ${O({ N_pF})}$. 
\begin{equation}
\label{delete}
{\mathcal{T}}_m=\{ {\mathbf{v}_{m,l}} \in {\mathbf{V_m}} \mid z_{m,l}=z_{i,j}, \forall ~ m \neq i , l\neq j \}
\end{equation}
With $\mathcal{T}_m$, the anchors are refreshed into a compact manner as ${\mathbf{A}_{c}}\!=\!\{a_m\!\mid\!{\mathcal{T}}_m\!\neq\!\emptyset\}$. 
We define LFTs ${\mathbf{T}\!=\!\{\mathcal{T}_m\!\mid\!{\mathcal{T}}_m\!\neq\! \emptyset\}}$ and ${\mathbf{V}_c^{e}\!=\!\{\mathbf{V}^e_m\!\mid\!{\mathcal{T}}_m\!\neq\!\emptyset\}}$. Up to this step, we have greatly removed the redundant candidates and increased the entropy of features by using $\mathbf{T}$ instead of ${{\mathbf{V}}}$. An illustration of $\mathbf{T}_{m^{\prime}}\in\mathbf{T}$ is shown in Fig.~\ref{fig:tensors},  where $m^{\prime}$ is the new index corresponding to $m$ after redundancy elimination.
\subsection{Hash-based fast duplication searching}
\label{lab:hbfds}
If the number of elements in ${\mathbf{T}_{m^{\prime}} \cap \mathbf{T}_{n^{\prime}} }$ is greater than ${k}$, we consider it as a suspected duplication, where ${k}$ is a small integer. As shown in Fig.~\ref{fig:tensors}, if select $k$=3,  $a^c_{m^{\prime}}$ and $a^{c}_{n^{\prime}}\in \mathbf{A_c}$ are the pairs of matched anchors as it has four vectors matched. The process of iterative matching ${\mathbf{T}_{m^{\prime}} \cap \mathbf{T}_{n^{\prime}} }$ can be given by 
\begin{equation}
\label{eq:binwisesearch}
\mathbf{M}=\{ (m^{\prime},n^{\prime}) \mid \lvert{\mathbf{T}_{m^{\prime}} \cap \mathbf{T}_{n^{\prime}}}\rvert\!\ge k,k\leq F, \forall m^{\prime}\neq n^{\prime}\}
\end{equation}
Assuming total $N_{a}$ anchors, the computational complexity of (\ref{eq:binwisesearch}) is ${O(F^2N_{a}^2)}$.  
The task of searching for duplications in (\ref{eq:binwisesearch}) is equivalent to iteratively matching among all of the $l$-element subsets\cite{Parque} of the tensors in $\mathbf{T}$, where $l\!\geq\!k$. However, this problem can be transformed by detecting duplication specifically among the k-element subsets of the tensors. It is based on the fact that ${\forall k<g}$, ${\mathbf{T}_{m^{\prime}}^{k,i} \in \mathbf{T}_{m^{\prime}}^k}$, ${i\in[0, C^k_F)}$, then ${\exists \mathbf{T}_{m^{\prime}}^{g,j}\in \mathbf{T}_{m^{\prime}}^g}$, ${j\in[0, C^j_F)}$, such that ${\mathbf{T}_{m^{\prime}}^{k,i}  \subset \mathbf{T}_{m^{\prime}}^{g,j}}$, where $\mathbf{T}_{{m^{\prime}}}^{k}$ is the $k$-element subsets of $\mathbf{T}_{m^{\prime}}$, and $\mathbf{T}_{{m^{\prime}}}^{k, i}$ denotes $i^{th}$ member of $\mathbf{T}_{{m^{\prime}}}^{k}$.  
Therefore, a duplication of ${\mathbf{T}_{m^{\prime}}^k}$ guarantees duplication of ${\mathbf{T}_{m^{\prime}}^g}$.
If the duration of duplication ${T_d}$ is sufficiently long, indicating that $N_{a}\!\gg\!F$, and typically ${k\!\leq3}$, the computational complexity can be reduced to ${O(N_{a}C_F^k)}$ by utilizing hash-based duplicate detection that matches the hashing of the k-element subsets. Considering $N_{a} \approx \kappa T_d+\gamma T_s$ and $N_p\approx \kappa T_s$, the proposed method above exhibits a linear time complexity of ${O(\kappa C^k_F T_d +\gamma C^k_FT_s+ \kappa FT_s})$ for processing (\ref{delete}) and (\ref{eq:binwisesearch}), where $\gamma$ represents the false collisions of $\mathbf{v}_{m,l}$ per second within (\ref{delete}), and typically with $\gamma<1$. 
Given that ${\mathbf{V}_c^{e}}$ contains the timestamps, we can readily obtain the specific time positions of duplication from it.

In order to perform an analysis of the above method, let us assume ${p}$ is the probability of a peak surviving in a noise attack. It generally is monotonic increasing with the signal-to-noise ratio (SNR). 
Considering the state ${E^a_{m^{\prime}}}\!=\!\{0,1\}$ and ${E^a_{n^{\prime}}}\!=\!\{0,1\}$ representing the survival of the anchors ${a^c_{m^{\prime}}}$ and ${a^c_{n^{\prime}}}$, if the entire speech is attacked by noise, the probability of both of the anchors can survive is given by ${p^a_{2}\!=\!P\left(E^a_{m^{\prime}}\!\land\! E^a_{n^{\prime}}\!=1\!\right)}\!=\!p^2$. Consequently, the probability of duplication that there are ${k}$ or more matched pairs of vector  is as follows:
\begin{equation}
\label{prob}
P_{d}\!\left(k\right)\!=\!\sum_{E^a_{m^{\prime}},E^a_{m^{\prime}}}P_{d}\left(k, E^a_{m^{\prime}}, E^a_{n^{\prime}}\right)\!=\!p^a_{2}P_{d}\left(k|E^a_{m^{\prime}}\land E^a_{n^{\prime}}\!=\!1\right) 
\end{equation}
Similar to the anchor pairs, the matched satellite pair ${(s_{m^{\prime},l}, s_{n^{\prime},l})}$ has three possible states, and the probability of each pair's state can be expressed as follows:
\begin{equation}
\begin{split}
State\ 2: p^s_{2}&=P(E^{s_{m^{\prime},l}}_{m^{\prime}} \land E^{s_{n^{\prime},l}}_{n^{\prime}}=1)=p^2\\
State\ 1: p^s_{1}&=P(E^{s_{m^{\prime},l}}_{m^{\prime}}\oplus E^{s_{n^{\prime},l}}_{n^{\prime}}=1)=2(1-p)p\\
State\ 0: p^s_{0}&=P(E^{s_{m^{\prime},l}}_{m^{\prime}} \lor E^{s_{n^{\prime},l}}_{n^{\prime}}=0)=(1-p)^2
\end{split}
\end{equation}
Assuming that the satellites around the anchors are IID, 
when absent of noise, 
all satellites survive, resulting in a total of F satellite pairs ${(s_{m^{\prime},l}, s_{n^{\prime},l})}$, all in state 2.  However, when noise is present, $\lvert{\mathbf{T}_{m^{\prime}} \cap \mathbf{T}_{n^{\prime}}}\rvert\!\ge j$, it implies that there are j satellite pairs in state 2, and the remaining $F-j$ satellite pairs are either in state 0 or state 1. Finally, (\ref{prob}) can be obtained by summing up the combinations of the probabilities associated with the states of the anchors and the satellite pairs.
\begin{equation}
\begin{split}
\label{prob2}
P_{d}\left(k\right)=p^a_{2}\sum_{j=k}^{F}C_{F}^{j}{(p^s_{2}})^{j}[{\sum_{i=0}^{F-j}C_{F-k}^{i}(p^s_{1})^{i}(p^s_{0})^{F-k-i}]}
\end{split}
\end{equation}
Finally, the probability of at least one pair of tensors existing in the speech, which represents the recall of detection, can be formulated as follows:
\begin{equation}
\label{exist}
P_{exist}=1-{(1-P_d\left(k\right))}^{{\mathbf{\mathbb{E}}({N_h})}}
\end{equation}
where ${\mathbb{E}(N_h)\!\approx p\kappa T_d}$. Typically, the duplication is easily found as long as ${\kappa}T_d$ is large enough as shown in Fig.\ref{fig:${k}$=3} 


\begin{figure}[ht]
     \begin{subfigure}[t]{0.24\textwidth}
         \includegraphics[width=\textwidth]{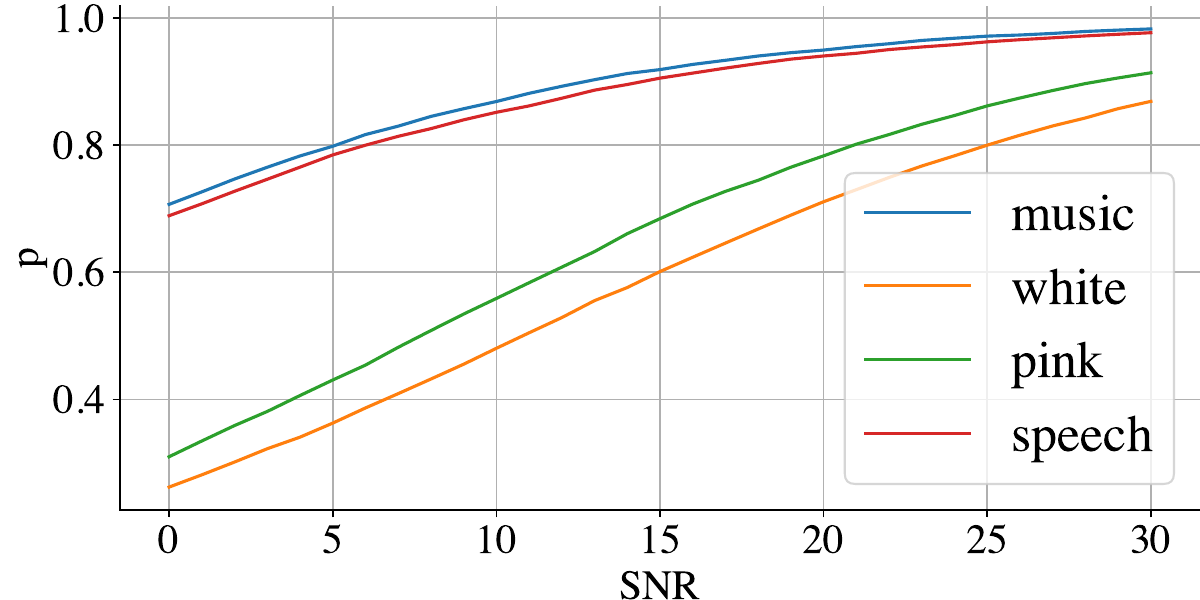}
         \caption{${p}$ vs ${SNR}$}
         \label{fig:snrvsq}
     \end{subfigure}
     \hspace{-2mm}
     \begin{subfigure}[t]{0.24\textwidth}
         \includegraphics[width=\textwidth]{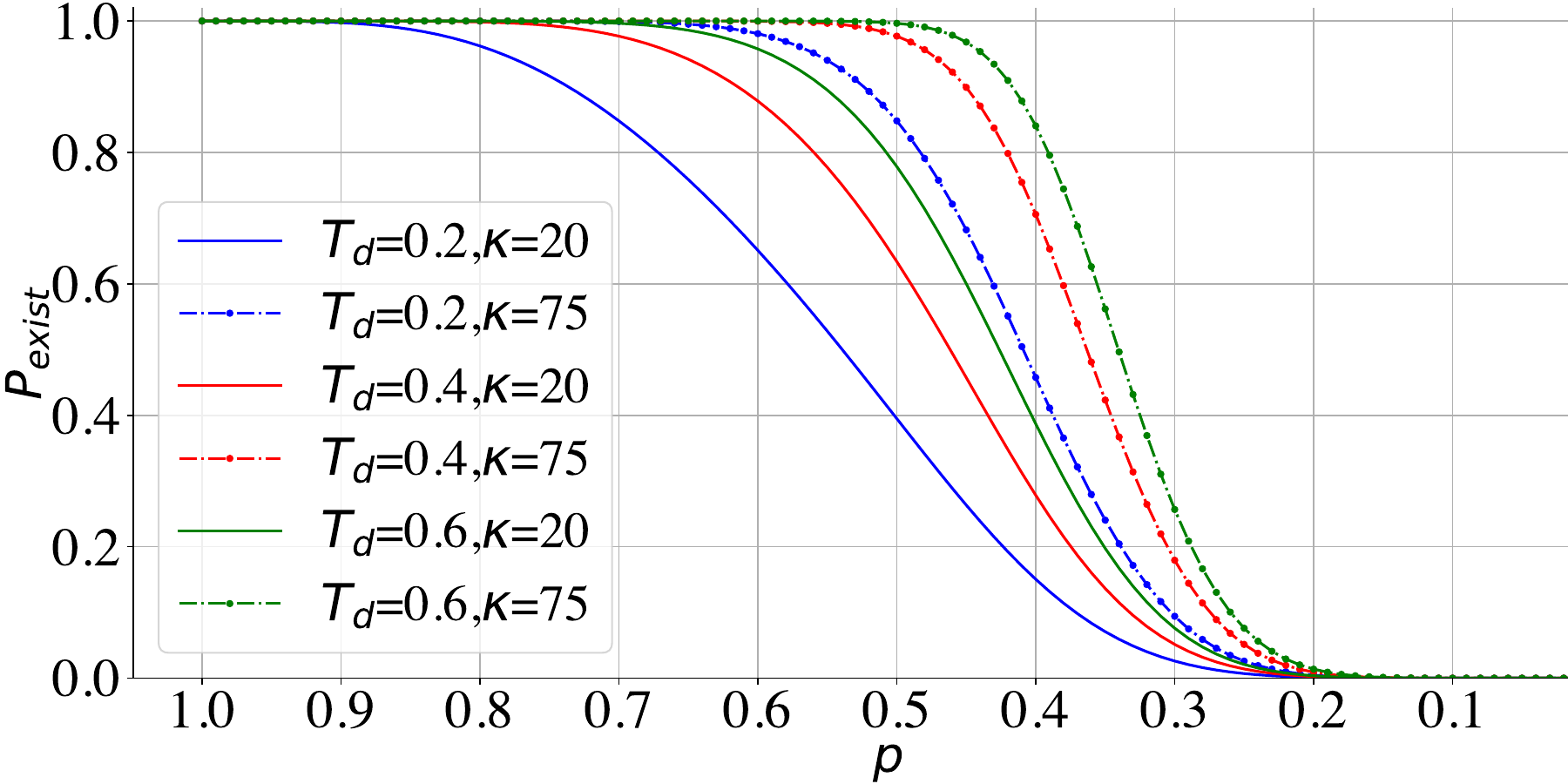}
         \caption{$k$=3, $F$=19, ${\kappa}$=25 and 75 }
         \label{fig:${k}$=3}
     \end{subfigure}
        \caption{(a) ${p}$ vs ${SNR}$ for various types of noise. (b) results of ${P_{exist}}$ vs ${p}$ applying with~(\ref{exist}) for different $T_d$ and ${\kappa}$.}
     \label{fig:curvekis3}
\end{figure}
\subsection{Local pitch distance analysis}
In order to mitigate residual accidental false alarms, the dynamic time-warping\cite{Senin2008DynamicTW} pitch distance factor (DTW-PDF) is proposed to assess the confidence of the duplication from the results based on  (\ref{eq:binwisesearch}). 
If we set threshold ${\theta}\in (0, 1]$, duplication finally is confirmed by
\begin{equation}
\mathbf{M}^{f}\!=\!\{(m^{\prime},\!n^{\prime})\!\mid\!\{D^{e}({\mathbf{P}}^D_{m^{\prime}},\!{\mathbf{P}}^D_{n^{\prime}})/{C_b}\!\leq\!{\theta},(m^{\prime},\!n^{\prime})\!\in \!\mathbf{M}\}
\end{equation}
where $C_b$ refers to the confidence boundary representing the maximum acceptable standard deviation for pitch estimation. The pitch slice is defined as ${{\mathbf{P}}^D_{m^{\prime}}=\mathbf{P}_s(m^{\prime}-D:m^{\prime}+D)}$. Pitch sequence $\mathbf{P}_s$ can be estimated using REAPER\cite{talkin2015reaper}\cite{Jouvet}. Due to potential inconsistencies in pitch estimation, the effective lengths of the ${{\mathbf{P}}^D_{m^{\prime}}}$ and ${{\mathbf{P}}^D_{m^{\prime}}}$ may not be necessarily aligned. To address this, the Euclidean distance-based dynamic time warping operator ${D^e}$ is employed to handle the situation.

\section{Performance Evaluation}
In this work, the TIMIT speech database\cite{TIMIT} was used to generate test copy-move datasets, which contain a total of 6300 sentences. Multiple sentences are randomly concatenated to a speech restricted between 12 and 60 seconds. For each speech recording, a segment of fixed duration is randomly selected and then randomly copied to another location within the same speech. The duration of duplication $T_d$ ranges from 0.2 to 1.0 seconds. The positive and negative samples are evenly distributed in the dataset. Precision and recall were used to measure the performance of the proposed method. We employ FFT size ${L}$=512 along with a hanning window of length $L_w$=512, and hop length H=64 for STFT. Pitch frequency ${{F_{0}}}$ is estimated by REAPER\cite{talkin2015reaper}. Parameters of ${\mathbf{B}_s=[3, max(8,\frac{2F_{0}L}{R}),15,1]}$, ${\mathbf{B}_d=[8,15]}$, $k$=3, $F$=19, $C_b$=10~(Hz), $D$=5, and threshold ${\theta}=0.5$. The target zone ${\mathbf{N}_m}\!=\!\{(i,j)\mid \!\lvert i-i^{'}\rvert \! \leq\!\frac{N_x}{2}, 0\!\leq\!j\!\leq\!\frac{L}{2}\}$ is a rectangle centered at ${a^c_m=(i^{'},j^{'})}$ and ${N_x}\!=\!40$. The experimental results show our method achieves 0.008 real-time factor on a 2.6 GHz Intel i7 core implemented by Python 3.8.

\begin{figure}
     \centering
     \includegraphics[width=0.5\textwidth]{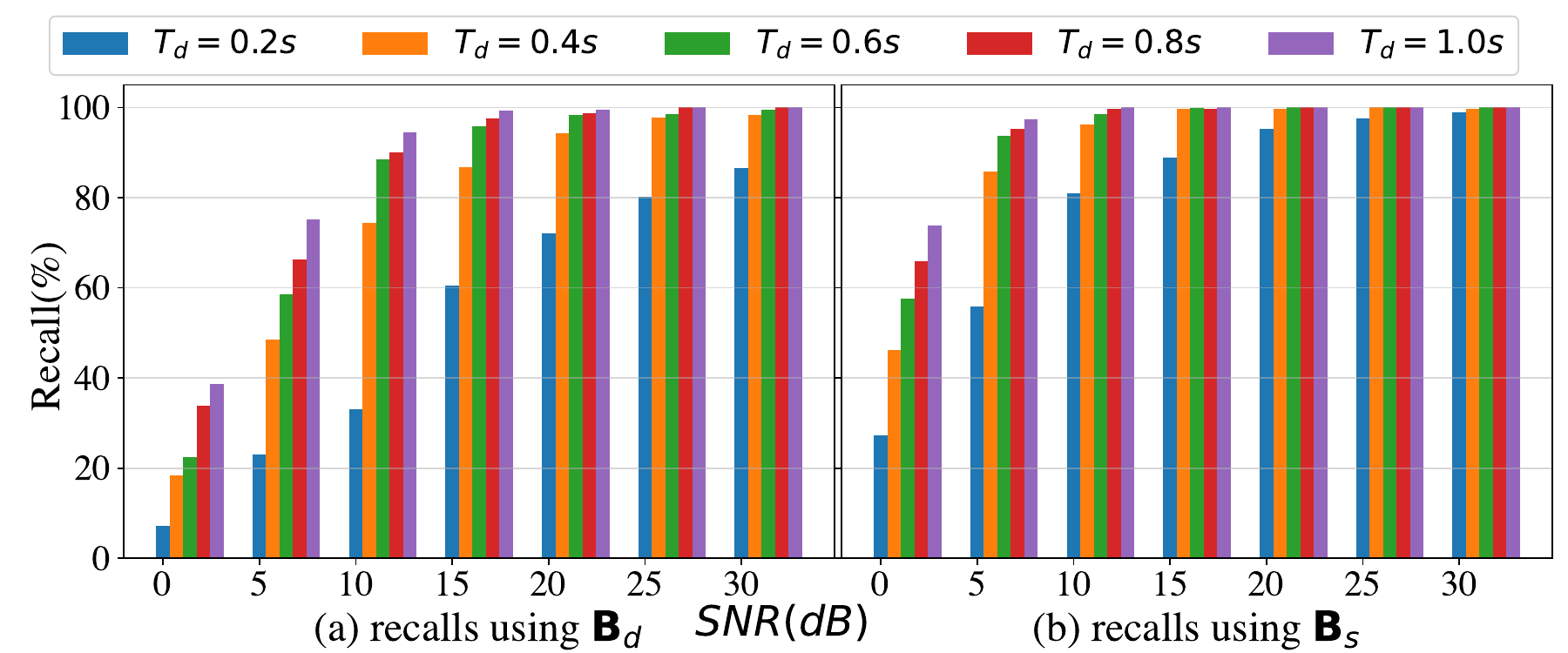}
        \caption{Performance for the speech under white noise attack at different SNR levels for various duplicate duration $T_d$. (a) and (b) recalls using ${\mathbf{B}_d}$ and ${\mathbf{B}_s}$ respectively. }
     \label{fig:curvekis4}
\end{figure}

Fig \ref{fig:${k}$=3} shows the effect of surviving probability $p$ on ${P_{exist}}$ for various choose of ${T_d}$. In connection with Fig \ref{fig:snrvsq}, under the same SNR, steady-state noise such as white noise has a greater impact on performance than transient noise. Therefore, we generate the copy-move forged speech datasets with additive white noise. The experimental results of our proposed method demonstrate comparable average precision, achieving 99.70\% and 99.68\% using DNB and SNB, respectively. Fig.~\ref{fig:curvekis4} shows the recall at various SNR levels for both cases using DNB and SNB. When selecting an SNB, the performance significantly exceeds its dense counterpart. By comparing Fig.~\ref{fig:curvekis4}a, Fig.~\ref{fig:curvekis4}b and Fig \ref{fig:${k}$=3}, the overall trend is consistent with formula~(\ref{exist}). When selecting a DNB, the peak distribution becomes sparser, leading to a reduced number of $\kappa$ and more closely aligning with the assumption of IID. In contrast, employing an SNB leads to an increased number of valid anchors $N_a$. Additionally, these peaks possess a more harmonically structural pattern and mutual dependency, along with a higher survival probability $p$ at a lower SNR. Consequently, this ultimately yields improved performance according to (\ref{exist}). Tab.~\ref{tab:perfoncodecs} shows our method also demonstrates robustness against compression degradation at very low bit rates, filtering, and resampling attacks.  

To fairly compare with other methods, five types of common post-processing operations \cite{Suzhaipin} are chosen to attack, including MP3 degradation, Gaussian noise with SNR=10dB and 20dB respectively, resampling, and filtering. LibriSpeech \cite{librspeech} and Chinspeech \cite{Suzhaipin} are used to generate test datasets as the same procedure in \cite{Suzhaipin}. Tab.\ref{tab:compareothers} presents the overall performance across five types of anti-forensics attacks, indicating that our method generally outperforms the methods proposed by Yan\cite{Yan2017} and Z. Yang's CQSS-299\cite{Suzhaipin}. Particularly, Yan's method or similar approaches are not competitive primarily due to their reliance on the assumption of between-utterance silence for speech segmentation, which is not suitable for real-world applications. 
\begin{table}[H]
\centering
\caption{Performance (\%) under different anti-forensics attacks using ${\mathbf{B}_s}$ for various ${T_d}$.} 
\label{tab:perfoncodecs}
\renewcommand{\arraystretch}{0.8}
\tabcolsep=0.15cm
\begin{tabular}{lcccccc}

\toprule
\multirow{2}*{Codecs} & & & {Recall}  & & & \multirow{2}{*}{Precision } \\
\cmidrule(lr){2-6} 
& {0.2}&  {0.4} &  {0.6 }&  {0.8} &  {1.0} \\
\midrule
                        PCM         &   99.85   &  100.0   &   100.0  &   100.0    &   100.0  &  99.33 \\ 
                        MP3(32k)    &   97.88    &   100.0      &   100.0   &    100.0   &   100.0  & 99.62 \\  
                        Opus(6k)    &   97.51    &   100.0    &   100.0     & 100.0    &   100.0  & 99.95\\
                        Lowpass\cite{Milic}   &   99.75    &   100.0    &   100.0     & 100.0    &   100.0  & 99.58\\
                        Resampling   &  99.78    &   100.0    &   100.0     & 100.0    &   100.0  & 99.68\\
\midrule
\bottomrule
\end{tabular}
\end{table}
\begin{table}[h]
\centering
\caption{Average Performance (\%) under five types of anti-forensics attacks.}
\label{tab:compareothers}
\renewcommand{\arraystretch}{0.8}
\tabcolsep=0.10cm
\begin{tabular}{lccccccccc}

\toprule
\multirow{2}*{Methods} & \multicolumn{3}{c}{LibriSpeech} & \multicolumn{3}{c}{ChinSpeech}  & \\
\cmidrule(lr){2-4} 
\cmidrule(lr){5-7} 
& {Yan} & {CQSS-299} & {Proposed} &{Yan} &  {CQSS-299} & {Proposed}  \\
\midrule
                        Precision  &  85.6  &  97.1    &    ${\mathbf{99.6}}$  &  88.8     &  99.5   & ${\mathbf{99.7}}$  \\ 
                        Recall     &   81.0    &   97.2   & ${\mathbf{97.8}}$  &  90.1   &  ${\mathbf{97.0}}$   & 96.9    \\  

\midrule
\bottomrule
\end{tabular}
\end{table}

\section{Conclusion}
In this letter, a novel CMF detection and location method is proposed based on comparing the local tensor features. Different from previous works, we truly achieve fully blind and continuous analysis by inspecting the T-F structure of a speech recording without relying on any speech priors and segmentation. Our theoretical analysis demonstrates it performs effective and efficient speech CMF detection and localization in linear complexity. Via extensive experiments and comparisons, we validate the effectiveness of the proposed method, which is robust against various post-processing attacks. 

\vfill\pagebreak

\nocite{*}
\bibliographystyle{IEEEtran}
\small\bibliography{IEEEabrv, ref} 

\end{document}